# Dirac node lines in a two-dimensional bipartite square lattice

**Bo Yang, Xiaoming Zhang, Mingwen Zhao***

School of Physics and State Key Laboratory of Crystal Materials, Shandong University, Jinan 250100, China
* zmw@sdu.edu.cn

## ABSTRACT

As a new type of quantum matter, Dirac node line (DNL) semimetals are currently attracting widespread interest in condensed matter physics and material science. The DNL featured by a closed line consisting of linear band crossings in the lattice momentum space are mostly predicted in three-dimensional materials. Here, we propose tight-binding (TB) models of $p_z/p_{x,y}$ or $p_z/s$ orbitals in a two-dimensional (2D) bipartite square lattice for the 2D version of DNL semimetals. The DNL states in these models are caused by the inversion of the bands with different symmetries and thus robust again spin-orbit coupling (SOC). By means of first-principles calculations, we demonstrate two candidate 2D materials of these models: $Be_2C$ and $BeH_2$ monolayers, which have Fermi circles centered at $\Gamma$ (0,0) and K (1/2, 1/2) points, respectively. The topological nontriviality is verified by the non-zero topological invariant and the edge states. This work opens an avenue for design of 2D DNL semimetals.

## Introduction

The topologically nontrivial materials, such as topological insulators (TIs) [1,2], have attracted broad interest. Different from normal insulators, TIs are bulk insulators with exotic gapless edge or surface states protected by bulk band topology. As new topologically nontrivial materials, topological semimetals are distinct from the normal metals in the touching points between the valence and conduction bands featured by gapless cone-like dispersion. According to the band touching (BT) points, topological semimetals can be classified into three categories: topological Dirac (TD) [3], topological Weyl (TW) [4], and Dirac node line (DNL) semimetals [6-12]. TD and TW semimetals have fourfold and twofold degenerate BT points near the Fermi level, respectively, along with fascinating surface states, such as surface Dirac cones and Fermi-arc states. Their low-energy bulk excitations are Dirac or Weyl fermions. Unlike the isolated BT points in TD and TW semimetals, the BT points in DNL semimetals form a fully closed line at the Fermi level in the momentum space. As an interesting consequence for the surface states, a flat band appears which may be half-filled when the surface is electrically neutral. Such surface states could be an interesting platform for strong correlation physics [13]. This type of DNL states has been predicted in several three-dimensional (3D) materials, such as carbon allotropes [5], antiperovskite $Cu_3$(Pd,Zn)N [6,7], $Ca_3P$ [8], LaN [9], photonic crystals [10], a superhoneycomb lattice [11], and pure alkali earth metals [12].

It is noteworthy that topological materials always have two-dimensional (2D) versions. For example, graphene is a prototypical TD semimetal, which exhibits Dirac points that are robust to the extent that spin-orbit (SO) interaction in carbon is weak. In the absence of SO interactions, the Dirac points in graphene are topologically protected by the combination of inversion and time-reversal symmetries. Young and Kane proposed a two-site tight-binding (TB) model for the Dirac semimetals including DNL phases [14]. The 2D DNL phases have also been predicted in a mixed lattice composed of honeycomb and kagome lattice [15] and MX (M=Pd, Pt; X=S, Se, Te) monolayer [16]. These interesting works extend the concept of DNL from 3D to 2D systems. Although 2D systems can be more fragile, and typically require substrates that can influence the behavior, they offer additional tunability not available in 3D systems. Compared with the 3D DNL semimetals, the 2D DNL semimetals are far from being understood and the realistic materials are scare. In this work, we propose a two-site TB model of a bipartite square lattice and demonstrate robust DNL states in this 2D system. The DNL states are due to the band inversion between valence and conduction bands with different symmetries and thus robust against spin-orbit coupling (SOC). By means of first-principles calculations, we predict two candidate 2D materials for this model, $BeH_2$ and $Be_2C$ which have Fermi circles centered at $\Gamma$ (0,0) and K (1/2, 1/2) points, respectively. The topological nontriviality is verified by the non-zero topological invariant and the edge states.

**Tight-Binding Model**

We now introduce a simple tight-binding model for 2D DNL semimetals. We begin with a bipartite square lattice composing of the two sublattices of $p_z$ and $p_{x,y}$ orbitals, as shown in Fig. 1(a)-(c). The $p_z$ orbitals constitute a square lattice in the x-y plane with a lattice parameter of $a$ (Fig. 1(a)), while the square lattice of the $p_{x,y}$ orbital has a lattice constant of $\sqrt{2}/2a$ (Fig. 1(b)). The nearest neighbor hoppings of the two sublattices are respectively $-t$ and $-t'$ (t>0, t'>0), and the second neighbor hoppings are omitted. As the two sublattices are mixed together in a manner shown in Fig. 1(c), the hopping between the two sublattices is prohibited due to the different symmetries of $p_z$ and $p_{xy}$ orbitals. Therefore, we can obtained the eigenvalues of the three bands with

$$E_1 = -2t\big(cos(k_x a) + cos(k_y a)\big)...(1)$$

$$E_{2,3} = \Delta \pm 4t' \left| sin\left(\frac{k_x a}{2}\right) \times sin(\frac{k_y a}{2})\right|...(2)$$

where $\Delta$ represents the on-site energy difference between two sublattices. $E_1$ band is formed by the $p_z$ orbitals, while $E_{2,3}$ bands exhibit the $p_{x,y}$ orbital features. At the K(1/2, 1/2) point, $E_1$ band has the maximum of 4t, while $E_3$ attains the minimum of $\Delta-4t'$. If (t + t') > $\Delta$/4, band inversion of between $E_1$ and $E_3$ will take place near the Fermi level, as shown in Fig. 1(d), leading to the formation of DNL centered at the K (1/2, 1/2) point, as shown in Fig. 1(e). We also calculated the parities of the electron wavefunctions of the two bands at the K point, and found that accompanied by the band inversion the parities change correspondingly, as shown in Fig. 1(d). The band inversion in the model is not related to spin-orbit coupling (SOC). We calculated the $Z_2$ topological invariant according to the strategy proposed by Fu et al. [17], and revealed a nonzero $Z_2 = 1$. This confirms that the band inversion at the K (1/2, 1/2) point results in the topologically nontrivial electronic structure of the system. It is noteworthy that $E_1$ and $E_3$ bands have different originations and symmetries, spin-orbit coupling (SOC) cannot open up a band gap at the crossing node line [18], and the DNL phases are therefore robust.

Beside the $p_z+p_{xy}$ bipartite square lattice, the square lattice of $p_z$ and s orbitals, as shown in Fig. 2, also exhibits the same features. In this case, the three TB bands are:

$$E_1 = -2t\big(cos(k_x a) + cos(k_y a)\big) \ ...(3)$$

$$E_{2,3} = \Delta \pm 4t' \left| cos\left(\frac{k_x a}{2}\right) \times cos\left(\frac{k_y a}{2}\right) \right| \ ...(4)$$

$E_1$ and $E_{2,3}$ are contributed by the $p_z$ and s orbitals, respectively. The minimum -4t of the $E_1$ band resides at the $\Gamma(0,0)$ point, while the $E_2$ band reaches the maximum of $\Delta+4t'$ at the $\Gamma(0,0)$ point. The band inversion at the $\Gamma(0,0)$ point takes place as (t + t') > -$\Delta$/4. Accompanied by the band inversion at the $\Gamma(0,0)$ point, the lattice becomes a DNL semimetal with topologically nontrivial electronic structure. The DNL phases are also robust against the SOC protected by the symmetry of the lattices.

## Candidate Materials

### Be$_2$C monolayer

Beryllium can form 3D crystal (space group Fm-3m) with carbon, where each Be binds to four C atoms, and each C connects to eight Be atoms, leading to a stiochiometry of Be$_2$C. We cleave a Be$_2$C (001) bilayer from the bulk crystal and optimize it without any symmetry constrain by using first-principles calculations. A planar Be$_2$C monolayer structure is finally obtained, as shown in Fig. 3(a). Compared with the bulk Be$_2$C crystal, the coordination numbers of the C and Be atom in the monolayer are reduced. Each C atom is coordinated by four Be atoms, while each Be connecting to two C atoms and four Be atoms. The $Be_2C$ monolayer possesses a bipartite square lattice with a lattice constant of 3.276 Å. The Be-C bond length 1.638 Å is shorter than that 1.876 Å in bulk $Be_2C$ crystal. The nearest Be-Be distance is 2.316 Å. This type of $Be_2C$ monolayer with planar tetracoordinate carbon has also been proposed as a local minimum structure by using the first-principles based particle swarm optimization (PSO) method [19].

The electronic band structures of the $Be_2C$ monolayer along the highly-symmetric directions (Γ-K-M-Γ) in the Brillouin zone are plotted in Fig. 3(b). It is interesting to see that at the Fermi level the valence band and conduction band meet at the points (P1 and P2) between Γ and K and between K and X, which is quite similar to the TB band structure shown in Fig. 1(d). The P1-K and K-P2 distances in the momentum space are 0.304 and 0.312 $Å^{-1}$, respectively. In the vicinity of the meeting points, the electronic bands exhibit linear dispersion relation along the Γ-K and K-M directions. The Fermi velocities ($v_F$) along the two directions calculated by the formula, $v_F = \frac{1}{\hbar}\frac{\partial E}{\partial k}$ are respectively $3.41 \times 10^6$ m/s and $3.48 \times 10^6$ m/s, both of which are larger than that in graphene $0.8 \times 10^6$ m/s. High carrier mobility can therefore be expectable in $Be_2C$ monolayer.

We then calculated the electronic band structures in the whole BZ using a two-dimension wave vector mesh and plotted them in the inset of Fig. 3(b). It is clear that the conduction band and valence band meet at a circle centered at the K point rather than isolated points, exhibiting features of DNL semimetals. The planar Fermi circle has a radius of about 0.60 $Å^{-1}$. These features are consistent with the TB model of $p_z/p_{x,y}$ orbitals.

The orbital contributions of the two bands (band I and II) near the Fermi level are quite obvious from the orbital-resolved electronic band structures, as shown in Fig. 3(c). Clearly, the two bands have different origins. Band I arises mainly from the in-plane $p_{x,y}$ orbitals of Be atoms, while band II is largely contributed by the $p_z$ orbitals of C atoms perpendicular to the monolayer. Therefore, $Be_2C$ monolayer can be regarded as a candidate material for the TB model of $p_z/p_{xy}$ bipartite square lattice.

The appearance of DNL states agrees with the topological prediction of $Z_2$ invariant calculated from the parity analysis. The spatial inversion symmetry of the $Be_2C$ monolayer allows us to calculate the $Z_2$ invariant from the parities of the electron wavefunctions at the time-reversal invariant momenta, without having to know about the global properties of the energy bands, as proposed by Fu et al [17]. In the square lattice of $Be_2C$ monolayer, there are four time-reversal invariant momenta: $\Gamma(0,0)$, $X(1/2,0)$, $X'(0,1/2)$, and $K(1/2,1/2)$.The $Z_2$ invariant $\nu$ is defined by

$$(-1)^\nu = \prod_i \delta_i \text{ with } \delta_i = \prod_{m=1}^{N} \xi_{2m}(\Gamma_i)$$

for 2N occupied bands. $\xi_{2m}(\Gamma_i) = \pm 1$ is the parity eigenvalue of the 2m-th occupied energy band at the time-reversal invariant momentum $\Gamma_i$. Our first-principles calculations showed that $\delta_i$ has the values of $-$, $-$, $-$, and $+$ at (0, 0), (1/2, 0), (0, 1/2), and (1/2, 1/2) time-reversal momenta, respectively, as shown Table I, leading to nonzero topological invariant $Z_2=1$. The topological nontriviality is closely related to the band inversion near the K point which leads to the change of parity, as shown in Fig. 3(b). These results are in good consistent with the TB model.

Another interesting issue is existence a flat band in the edge states of the 2D DNL semimetal. Using the recursive method [20], we constructed the edge Green's function of the semi-infinite $Be_2C$ lattice from maximally localized Wannier functions (MLWFs). The local density of states (LDOS) of the edge calculated from the MLWFs is plotted in Fig. 3(e). A nearly flat band appears between two crossing points, similar to the cases of 3D DNL semimetals. The flat band may be half-filled when the edge is electrically neutral, offering an interesting platform for strong correlation physics [13]. The appearance of flat band confirms the topological nontriviality of the $Be_2C$ monolayer.

**$BeH_2$ monolayer**

Beryllium hydride ($BeH_2$) monolayer has been proposed as a candidate material for superconductivity [21-23]. The atomic arrangement in $BeH_2$ monolayer is similar to the beryllium hydride plane of $LiBeH_3$ crystal [24]. Each Be atom is coordinated by four hydrogen atoms, and each hydrogen atom connects to two Be atoms, forming a planar square lattice, as shown in Fig. 4(a). The lattice constant of BeH$_2$ monolayer obtained from first-principles calculations is about 2.953 Å. The Be-H bond length is 1.476 Å, and the Be-Be distance is 2.088 Å.

The electronic band structures of BeH$_2$ monolayer calculated from first-principles calculations are plotted in Fig. 4(b). The band inversion near the $\Gamma(0,0)$ point is quite obvious, resulting in DNL centered at the $\Gamma(0,0)$ point, as shown in Fig. xx. The radius of the Fermi circle is 0.76 Å$^{-1}$. The Fermi velocities along the M-$\Gamma$ and $\Gamma$-K directions are 3.69×10$^6$m/s and 3.36×10$^6$m/s, respectively. From the orbital-resolved electronic band structures shown in Fig. 4(c), we can clearly see that the two bands are mainly contributed by the p$_z$ orbitals of Be and the s orbitals of H, respectively. These features are in good agreement with p$_z$/s TB model for DNL semimetals. The topologically nontrivial electronic structure of the BeH$_2$ monolayer is also confirmed by the nonzero Z$_2$ topological invariant, as shown in Table I. We also calculated the electronic structure of the BeH$_2$ monolayer by taking the SOC intro account and found that the DNL states are robust against SOC.

## Discussion

The realistic materials for the TB mode proposed in this work are challenging. Although both Be$_2$C [19] and BeH$_2$ [21-23] monolayer have been investigated theoretically in pervious works, experimental realization has not yet been reported. From energetic points of view, Be$_2$C monolayer is less stable than bulk crystal by about 0.635 eV/atom. However, such an energy difference is smaller than that between silicene and bulk Si crystal (0.83 eV/atom), which implies the plausibility of this monolayer. We also checked the dynamic stability of the Be$_2$C monolayer from the phonon spectrum calculated from first-principles. The phonon spectrum is free from imaginary frequency modes, as shown in Fig. 5(a), even in the long wave length region. This implies that the Be$_2$C monolayer once synthesized is dynamically stable. For the BeH$_2$ monolayer, there is no bulk counterpart for energetic comparison. The formation energy of BeH$_2$ monolayer with respect to bulk Be crystal and H$_2$ molecule is about 0.45 eV/atom, suggesting its energetic disadvantage. The phonon spectrum of free-standing $Be_2H$ monolayer has imaginary frequency modes, implying that it is unstable without the support of substrate.

We checked whether the topological properties of the $Be_2C$ monolayer can remain on substrates or not. We employed a hexagonal BN substrate which is expected to have a weak van der Waals interfacial interaction to support the monolayers, as shown in Fig. 5(b). The optimized distances between the supported $Be_2C$ monolayer and BN substrate is respectively about 3.406 Å, indicating the weak interaction between them. Electronic structure calculations show that the DNL states remain intact, as shown in Fig. 5(c). These results demonstrate the feasibility of attaining the DNL states of the $Be_2C$ monolayer on a substrate. For the $BeH_2$ monolayer, however, the structural fluctuation easily occurs even on the BN substrate, which opens a trivial band gap in the $BeH_2$ monolayer. This makes the realization of DNL states in $BeH_2$ monolayer more challenging.

## Conclusion

In summary, we propose a simple tight-binding model for two-dimensional Dirac node line semimetals. We demonstrate robust DNL states in a bipartite square lattice of $p_z$ and $p_{x,y}$ (or s) orbitals, which arise from the band inversion and weak spin-orbit coupling. By means of first-principles calculations, we predict two candidate 2D materials for this model: $Be_2C$ and $BeH_2$ monolayers which have Fermi circles centered at $\Gamma$ (0,0) and K (1/2, 1/2) points, respectively. The topological nontriviality is verified by the non-zero topological invariant and the edge states. This work opens up an avenue for achieving DNL states in 2D materials.

## First-Principles Calculations

Our first-principles calculations were performed within the framework of density functional theory (DFT), as implemented in the Vienna ab initio simulation package (VASP) [25]. The Kohn-Sham electron wavefunctions were expanded using the plane-wave functions with an energy cutoff of 520eV for the plane-wave expansion. The electron-ion interaction was described by projector-augmented wave (PAW) potential [26].The generalized gradient approximation (GGA) [27] in the form of Perdew-Burke-Ernzerhof (PBE) [28]was adopted for the exchange-correlation functional. Two-dimensional periodic boundary conditions were employed in the x-y plane, while a vacuum space up to 20 Å was applied along the z-direction to exclude the interaction between neighboring images. The Brillouin zone (BZ) was sampled using a $11 \times 11 \times 1$k-point grid [29].The atomic coordinates were fully relaxed using a conjugate gradient (CG) scheme without any symmetry restrictions until the maximum force on each atom was smaller than 0.01 eV /Å. The phonon spectra were calculated by using the Phononpy code [30] interfaced with VASP.

## Acknowledgement

This work is supported by the National Natural Science Foundation of China (No. 21433006), the 111 project (No. B13029), the National Key Research and Development Program of China grant 2016YFA0301200, and the National Super Computing Centre in Jinan.

## Author Contributions

B. Y. performed first-principles calculations and analyzed the data. X.Z. calculated the edge states and the tight-binding bands. M.Z conceived the research, proposed the tight-binding method and wrote the paper.

**Table I**

The parities of electron wavefunctions at time-reversal invariant momenta: $\Gamma(0,0)$, X(1/2,0), X'(0,1/2), and K(1/2,1/2).

(a) Be$_2$C monolayer

|              | PARITIES |   |   |   | $\delta$ |
|--------------|----|----|----|----|----|
| $\Gamma(0,0)$ | +  | −  | −  | −  | -1 |
| M(1/2,0)     | +  | −  | −  | −  | -1 |
| M'(0,1/2)    | +  | −  | −  | −  | -1 |
| K(1/2,1/2)   | +  | −  | −  | +  | 1  |

(b) BeH$_2$ monolayer

|              | PARITIES |   | $\Delta$ |
|--------------|----|----|----|
| $\Gamma(0,0)$ | +  | −  | -1 |
| M(1/2,0)     | −  | +  | -1 |
| M'(0,1/2)    | −  | +  | -1 |
| K(1/2,1/2)   | −  | −  | +1 |

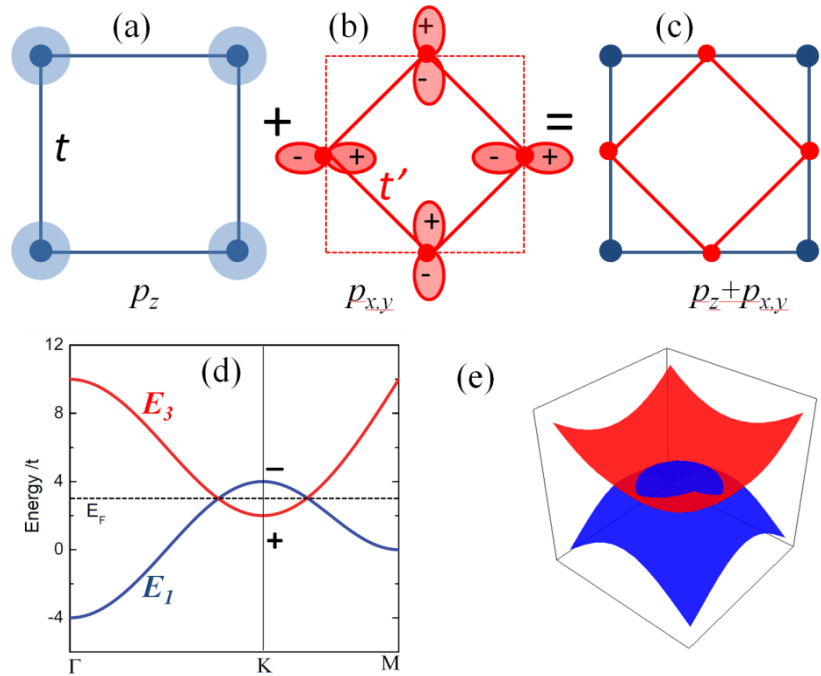

Fig. 1 Tight-binding model of DNL in a square lattice of $p_z$ and $p_{x,y}$ orbitals. (a) $p_z$-orbital sublattice; (b) $p_{x,y}$-orbital sublattice; (c) $p_z+p_{x,y}$ orbitals lattice; (d) tight-binding band inversion near the Fermi level, signs "+" and "-" denote the parity of the electron wavefunctions at the K (1/2,1/2) point d) DNL centered at the $\Gamma$ (1/2,1/2) point. The data were obtained by setting t' = 2t, $\Delta$ = 10t.

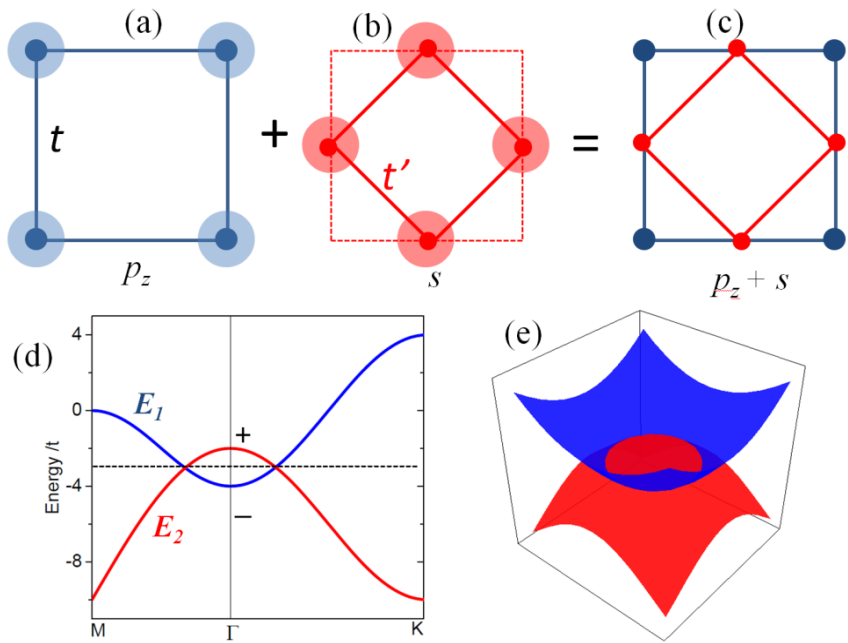

Fig. 2(a)-(c) schematic representation of a square lattice of $p_z$ and s orbitals. (a) $p_z$-orbital sublattice; (b) s-orbital sublattice; (c) $p_z$+s orbitals lattice; (d) tight-binding band inversion near the Fermi level, signs "+" and "-" denote the parity of the electron wavefunctions at the $\Gamma$ (0,0) point d) DNL centered at the $\Gamma$ (0,0) point. The data were obtained by setting t' = 2t, $\Delta$ = -10t.

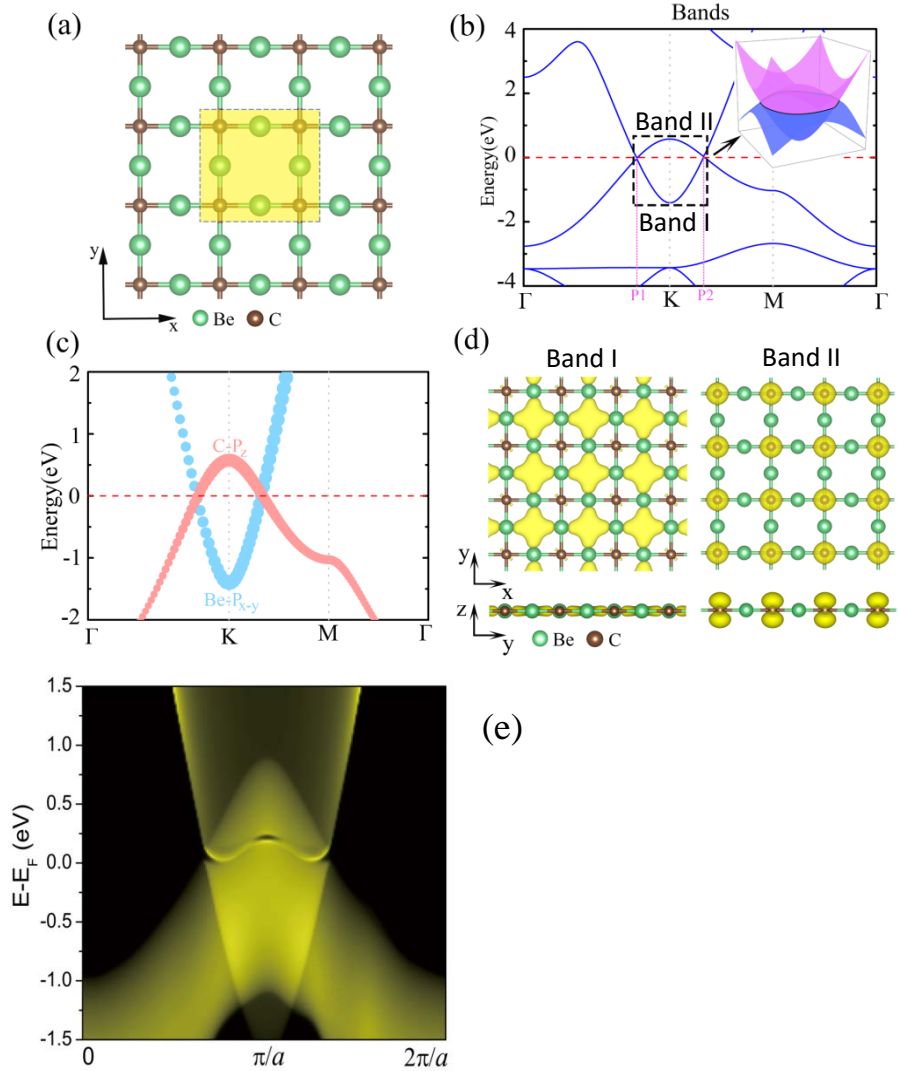

Fig. 3 (a) The square lattice and (b) electronic band structure of Be₂C monolayer. The DNL is shown in the inset of (b). The energy at the Fermi level is set to zero. (c) The orbital-resolved band structure of the Be₂C monolayer. (d) The electron density isosurfaces of the electron wavefunctions of band I and band II. (e) The local electron density of states (LDOS) of the edge calculated from localized Wannier functions (MLWFs).

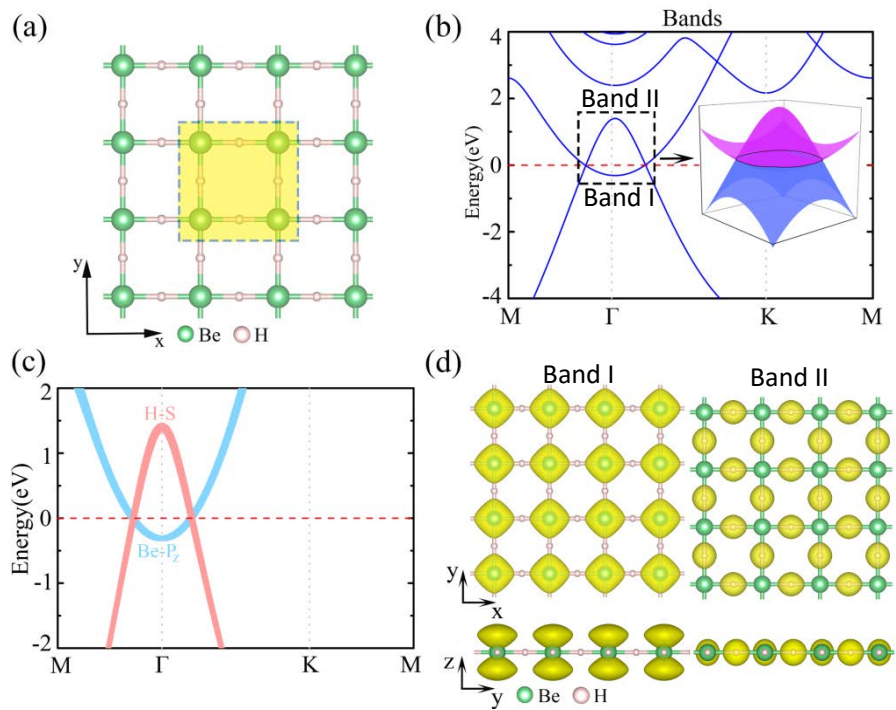

Fig. 4 (a) The bipartite square lattice of Be$_2$C monolayer. (b) Electronic band structure and (c) orbital-resolved band structure of Be$_2$H monolayer. The energy at the Fermi level is set to zero. (d) The electron density isosurfaces of the electron wavefunctions of band I and band II.

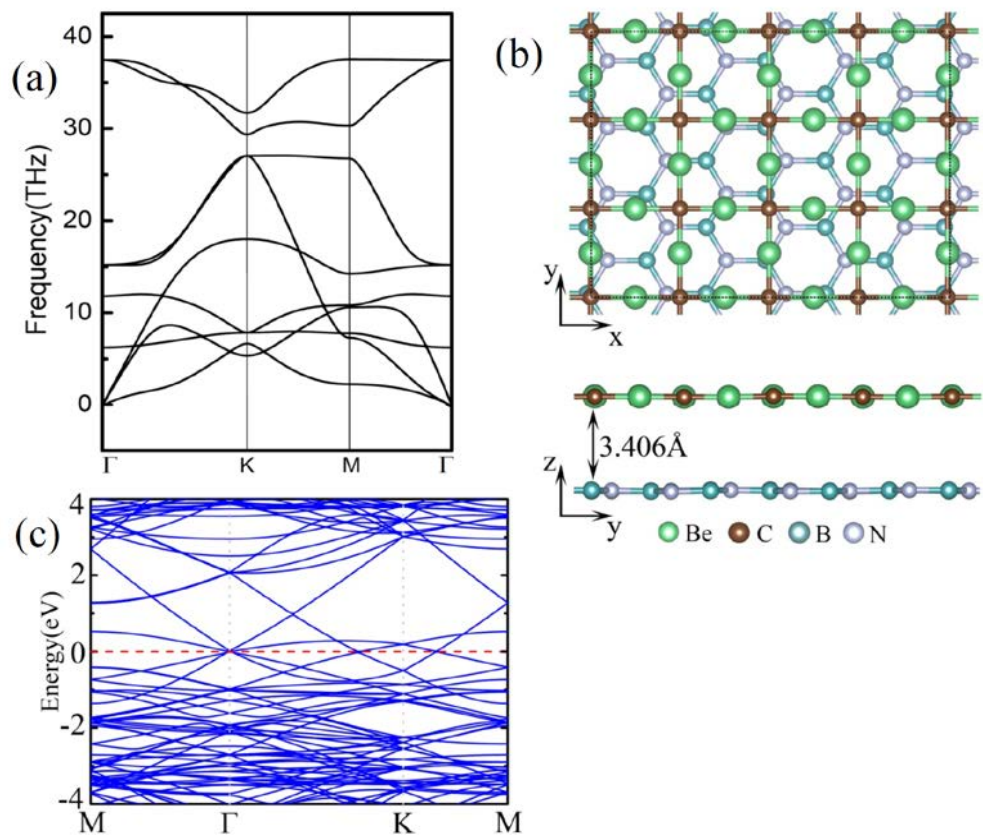

Fig. 5 (a) The phonon spectrum of Be$_2$C monolayer. (b) Atomic configuration and (c) electronic band structure of a Be$_2$C monolayer on a BN substrate. The energy at the Fermi level was set to zero.